\documentclass[letterpaper]{article} 
\usepackage{aaai24}  
\usepackage{times}  
\usepackage{helvet}  
\usepackage{courier}  
\usepackage[hyphens]{url}  
\usepackage{graphicx} 
\usepackage{natbib}  
\usepackage{caption} 
\usepackage{algorithm}
\usepackage{algorithmic}
\usepackage{tabularray}
\usepackage{graphicx}
\usepackage{subfig}
\usepackage{dsfont}
\usepackage{newfloat}
\usepackage{listings}
\DeclareCaptionStyle{ruled}{labelfont=normalfont,labelsep=colon,strut=off} 
\floatstyle{ruled}
\newfloat{listing}{tb}{lst}{}
\floatname{listing}{Listing}

\usepackage{color}

\nocopyright

\title{Decoding AI's Nudge: 
A Unified Framework to Predict Human Behavior in AI-assisted Decision Making
}

\author {
    Zhuoyan Li,
    Zhuoran Lu,
    Ming Yin
}
\affiliations {
    Purdue University, USA \\
    li4178@purdue.edu, lu800@purdue.edu, mingyin@purdue.edu
}

\usepackage{amsmath}

\usepackage{bm}
\usepackage{bibentry}

\begin{document}

\maketitle

\begin{abstract}

With the rapid development of AI-based decision aids, 
different forms of AI assistance have been increasingly integrated into the human decision making processes. 
To best support humans in decision making, it is essential to quantitatively understand how diverse forms of AI assistance influence humans' decision making behavior. To this end, much of the current research focuses on the end-to-end prediction of human behavior using ``black-box'' models, often lacking interpretations of the nuanced ways in which AI assistance impacts the human decision making process. 
Meanwhile, methods that prioritize the interpretability of human behavior predictions are often tailored for one specific form of AI assistance, making adaptations to other forms of assistance difficult. 
In this paper, we propose a computational framework that can provide an interpretable characterization of the influence of different forms of AI assistance on decision makers in AI-assisted decision making.  By conceptualizing  AI assistance as the ``{\em nudge}'' in human decision making processes, our approach centers around modelling how different forms of AI assistance modify humans' strategy in weighing different information in making their decisions. Evaluations on behavior data collected from real human decision makers 
show that the proposed framework outperforms various
baselines in accurately predicting human behavior in AI-assisted decision making. Based on the proposed framework, we further provide insights into how individuals with different cognitive styles are nudged by AI assistance differently.
\end{abstract}

\section{Introduction}
As AI technology advances, AI models are increasingly integrated into the human decision making process  spanning various domains from healthcare to finance. This has created a new paradigm of human-AI collaboration---Given a decision making task, AI provides assistance to humans while humans make the final decisions. To fully unlock the potential of AI-based decision aids in enhancing human decision making, a growing line of research has been developed in the human-computer interaction community to develop different forms of AI assistance to better support decision makers~\cite{Lai2023TowardsAS}. Each form of AI assistance shows different impacts on human decision makers.  
For instance, the most intuitive type of AI assistance is to directly provide the decision maker with an AI model's decision recommendation on a task~\cite{passi2022overreliance,wang2023effects}, though it is found that such assistance sometimes foster a degree of over-reliance on the AI recommendation~\cite{Ma2023WhoSI}. In contrast, the delayed recommendation paradigm,  another form of AI assistance where humans are first required to deliberate on a task before receiving the AI model's decision recommendation, has the potential to mitigate this over-reliance, but possibly at the cost of increased under-reliance~\cite{Buccinca2021ToTO, Fogliato2022WhoGF}. 
Thus, to best utilize diverse forms of AI assistance and to determine when and how to present the most suitable type of AI assistance to humans, it is critical to quantitatively understand and predict how these AI assistance influences humans on different decision making tasks.

A few existing studies have worked on modeling and predicting human behavior in AI-assisted decision making~\cite{kumar2021explaining,Bansal2021IsTM}, but they come with several limitations. For example, some research focuses on predicting humans' interaction with AI assistance in an end-to-end manner using black-box models~\cite{subrahmanian2017predicting}. While these methods can be effortlessly adapted to various forms of AI assistance,  the black-box nature of the model makes it challenging to unpack the cognitive mechanisms driving humans' decision behavior under AI influence.  Meanwhile, other studies that aim for interpretability propose computational models based on 
economics or psychology theories. For instance, \citeauthor{wang2022will}~\shortcite{wang2022will} employed the Cumulative Prospect Theory (CPT) to understand how humans decide whether to adopt AI recommendations by analyzing the utility and cost of different decision options. \citeauthor{kumar2021explaining}~\shortcite{kumar2021explaining}
captured humans' metacognitive processes of deciding when to rely on themselves and when to solicit AI assistance using  
the metacognitive bandits model.
However, these models 
are typically tailored for one specific form of AI assistance, making generalizations to other forms of AI assistance a challenging task that requires significant methodology adaptation.
 
The absence of a unified computational framework to quantitatively characterize how diverse forms of AI assistance influence human decision making processes in an interpretable way impedes the further intelligent utilization of AI assistance. As such, decision makers often have to interact with the default forms of AI assistance instead of benefiting from personalized and intelligent AI assistance that can best support them. 
Therefore, in this study, we aim to bridge this gap by proposing such a computational framework\footnote{In this study, we narrow down the scope of our framework to model the influence of AI assistance on human decision makers on each individual decision making task. That is, we do not consider the sequential or temporal influence of AI assistance on humans in a sequence of decision making tasks. 
}.

Specifically, inspired by \citeauthor{callaway2022optimal}~\shortcite{callaway2022optimal} that explores the designs of optimal nudges for cognitively bounded agents, we conceptualize the AI assistance as a ``{\em nudge}'' to the human decision making process, which would modify how humans weigh different information in making their decisions. 
Therefore, in our framework, we first establish an independent decision model that reflects how humans form their independent decisions without any AI assistance. We then model the nudge of AI assistance to humans as the alterations to their decision models.
To evaluate the performance of the proposed framework, we collect data on real human subjects' decisions in AI-assisted diabetes prediction tasks with the aids of three common types of AI assistance through a randomized experiment. By fitting various computational models to the behavior dataset collected, we find that our proposed framework consistently outperforms other baseline models in accurately predicting the human decision behavior under different forms of AI assistance. Furthermore, 
the proposed framework demonstrates robust performance in accurately predicting human behavior in AI-assisted decision making even with limited training data. Lastly, through a detailed analysis of the nudging effects of AI assistance identified by our framework, we offer quantitative insights into how individuals with different cognitive styles are nudged by AI assistance differently. For instance, we observed that 
AI explanations appear to show a larger effect in redirecting the attention of intuitive
decision makers than reflective decision makers.

\section{Related Work}
\noindent\textbf{Empirical Studies in AI-Assisted 
Decision Making.} The increased usage of decision aids driven by AI models has inspired a line of experimental studies that identify different forms of AI assistance to enhance human-AI collaboration in decision making~\cite{Lai2023TowardsAS}.  By surveying the literature related to AI-assisted decision making in the ACM Conference on Human Factors in Computing Systems, ACM Conference on Computer-supported Cooperative Work and
Social Computing, ACM Conference on Fairness, Accountability, and Transparency, and ACM Conference on Intelligent User Interfaces from 2018 to 2021, we identify three common types of AI assistance:
\begin{enumerate}
    \item \emph{Immediate assistance}: The AI model's decision recommendation on the decision making task and other indicators of the recommendation are provided to decision makers upfront. Typical indicators of the AI recommendation include the AI model's accuracy~\cite{Lai2020WhyI}, explanations of the AI recommendation~\cite{PoursabziSangdeh2018ManipulatingAM,Cheng2019ExplainingDA,SmithRenner2020NoEW,Liu2021UnderstandingTE,Tsai2021ExploringAP, Bansal2020DoesTW,Zhang2020EffectOC}, and confidence levels of the recommendation~\cite{Green2019ThePA, Guo2019VisualizingUA,Zhang2020EffectOC,Levy2021AssessingTI},  These indicators may help decision makers gauge the credibility of AI recommendation and calibrate their trust in AI. Since various indicators of the AI recommendation serve similar purposes, aligning with prior research~\cite{tejeda2022ai,wang2022will}, 
    we focus on modeling how immediate assistance influences human decision makers when the model's prediction confidence is used as the indicator in this study.

    \item \emph{Delayed recommendation}~\cite{Park2019ASA,GrgicHlaca2019HumanDM,Lu2021HumanRO,Buccinca2021ToTO,Fogliato2022WhoGF,Ma2023WhoSI}: Humans need to first make an initial decision on the task before the AI model's decision recommendation is revealed to them; this type of AI assistance forces humans to  engage more thoughtfully with the AI recommendation.

    \item \emph{Explanation only}~\cite{Lucic2019WhyDM, Alqaraawi2020EvaluatingSM,Rader2018ExplanationsAM,Schuff2022HumanIO,Berkel2021EffectOI}: Only the AI model's decision explanation but not its decision recommendation is provided to decision makers. The explanation often points out important features of the task that contribute the most to the AI model's unrevealed decision recommendation, aiming to highlight information that AI believes as highly relevant for decision making. 
  
\end{enumerate}

For more details of the literature review, please see the supplementary material. In this study, we focus on 
building a computational framework to characterize how different forms of AI assistance, such as the three types identified above, impact humans in AI-assisted decision making.

\vspace{2pt}
\noindent\textbf{Modeling Human Behavior in AI-assisted Decision Making.} Most recently, there has been a surge of interest among researchers in computationally modeling human behavior in AI-assisted decision making~\cite{Bansal2021IsTM,kumar2021explaining,tejeda2022ai, Pynadath2019AMM, Li_Lu_Yin_2023,lu2023strategic}.
Many of these studies build their models on economics frameworks~\cite{wang2022will}, which explain human decision making behavior under uncertainty, or on psychological frameworks that describe the relationship between human trust and reliance on automated systems~\cite{ajenaghughrure2019predictive,Li_Lu_Yin_2023}.
However, most existing works are either tailored to one specific form of AI assistance or lack interpretations of how AI assistance influences human decision making processes.
Inspired by the recent research in computationally modeling the effects of nudges~\cite{callaway2022optimal}, we take a different approach in this paper and build a framework to characterize diverse forms of AI assistance as nudges in the human decision making process.

\section{Methods}

\subsection{Problem Formulation}
We now formally describe the AI-assisted decision making scenario in this study. Suppose a decision task can be characterized by an $n$-dimensional feature $\bm{x} \in \mathcal{R}^{n}$, and $y$ is the correct decision to make in this task. In this study, we focus on decision making tasks with binary choices of decisions, i.e., $y \in \{0,1\}$, and each feature $x_i$ of the task $\bm{x}$ is normalized to fall within the interval of $[0,1]$. We use $\mathcal{M}(\bm{x};\bm{w}_m)$ to denote the AI model's output on the decision task ($\bm{w}_m$ are model parameters), and it is within the range of $[0,1]$. Given $\mathcal{M}(\bm{x};\bm{w}_m)$, the AI model can provide a binary decision recommendation to the human decision maker (DM), i.e., $y^m = \mathds{1}( \mathcal{M}(\bm{x};\bm{w}_m) > 0.5 )$. 
 The AI model's confidence in this recommended decision is $c^m = \max \{\mathcal{M}(\bm{x};\bm{w}_m), 1-\mathcal{M}(\bm{x};\bm{w}_m)\}$. Following explainable AI methods like LIME~\cite{ribeiro2016should} and SHAP~\cite{lundberg2017unified}, the AI model could also provide some ``explanations'' of its decision recommendation, $\bm{e} = \mathcal{E}(\mathcal{M}(\bm{x};\bm{w}_m)), \bm{e} \in \mathcal{R}^{n}$, by  
 highlighting the ``important'' features that contribute the most to the decision recommendation. Here, $e_i \in \{0,1\}$, where $e_i = 1$ means the feature $x_i$ is highlighted as important, while $e_i = 0$ means the feature $x_i$ is not highlighted.  In addition, we assume that the human DM also independently forms their own judgment of the decision task, which is characterized by the function $\mathcal{H}(\bm{x};\bm{w}_h)$ whose output is in the range of $[0,1]$.  Thus, $y^h = \mathds{1}( \mathcal{H}(\bm{x};\bm{w}_h) > 0.5 )$ represents the human DM's independent binary decision.

We consider the setting where the human DM is asked to complete a set of $T$ decision tasks with the help of the AI model. For each task $t$ $(1\leq t \leq T)$, the human DM is given the feature vector $\bm{x}^t$ and the AI assistance. As discussed previously, we focus on studying the following three forms of AI assistance: 
\begin{itemize}
\label{ai_treament}
    \item \emph{Immediate assistance:} The AI model's binary decision recommendation $y^{m,t}$ and its confidence $c^{m,t}$ are immediately provided to the DM along with the task $\bm{x}^t$.

    \item \emph{Delayed recommendation:} The DM is required to first make an initial independent decision $y^{h,t}$ on the task. After that,  
    the AI model's binary decision recommendation $y^{m,t}$ will be revealed to the DM.
      
    \item \emph{Explanation only:} The DM is only provided with the AI model's explanation $\bm{e}^t$, which highlights the important features of the task that contributes the most to the AI model's unrevealed decision recommendation.

\end{itemize}
The DM's independent judgement on the task is $y^{h,t}$---this is observed as the DM's initial decision when AI assistance comes in the form of 
\emph{delayed recommendation}, but is unobserved (thus requires inference) when AI assistance comes in the other two forms. 
  
Given both their own judgement and the AI assistance, the DM then makes a final decision $\hat{y}^t$ on the task. The goal of our study is to quantitatively characterize 
how the DM is ``nudged'' by different forms of AI assistance in making their final decision on each task.

\subsection{Model Decision Makers' Independent Judgement}
 
To characterize how AI assistance nudges human DMs in AI-assisted decision making, 
it is necessary to first understand how human DMs form their independent judgement {\em without} being nudged by AI. That is, we need to quantify human DMs' independent decision model $\mathcal{H}(\bm{x};\bm{w}_h)$.  Since each DM may have their own unique independent decision making model with different model parameters $\bm{w}_h$, given a training dataset of the DM's independent decisions $ \mathcal{D} = \{\bm{x}_i, y_i^{h}\}_{i=1}^{N}$, we adopt a Bayesian approach and set out to learn from the training dataset the posterior {\em distribution} of model parameters for a population of diverse DMs, i.e., $\mathcal{P}(\bm{w}_h|\mathcal{D})$,  instead of learning a point estimate. As directly computing this posterior $\mathcal{P}(\bm{w}_h|\mathcal{D})$ is intractable, we leverage variational inference to approximate it using the parameterized distribution $q_{\phi}(\bm{w}_h) = \mathcal{N}(\bm{w}_h;\bm{\mu}_{\phi},\bm{\Sigma}_{\phi})$ and minimize the KL divergence between $q_{\phi}(\bm{w}_h)$ and $\mathcal{P}(\bm{w}_h|\mathcal{D})$:
\begin{small}
\begin{equation}
\begin{split}    
    &\mbox{KL}(q_{\phi}(\bm{w}_h)\|\mathcal{P}(\bm{w}_h|\mathcal{D})) = \int_{\bm{w}_h} q_{\phi}(\bm{w}_h) \log \frac{ q_{\phi}(\bm{w}_h)}{\mathcal{P}(\bm{w}_h|\mathcal{D})} \mathrm{d}\bm{w}_h  \\
    & =\int_{\bm{w}_h} q_{\phi}(\bm{w}_h) ( \log\frac{ q_{\phi}(\bm{w}_h)}{ \mathcal{P}(\bm{w}_h) } - \log \mathcal{P}(\mathcal{D}|\bm{w}_h) + \log \mathcal{P}(\mathcal{D}) )\mathrm{d}\bm{\bm{w}_h} \\
    & =  \mbox{KL}(q_{\phi}(\bm{w}_h)\|\mathcal{P}(\bm{w}_h)) -  \mathrm{E}_{q_{\phi}(\bm{w}_h)}[\log \mathcal{P}(\mathcal{D}|\bm{\bm{w}_h}) -\log \mathcal{P}(\mathcal{D})]
\end{split}
\end{equation}
\end{small}
\noindent where $\mathcal{P}(\bm{w}_h)$ is the prior distribution of $\bm{w}_h$ and  $\mathcal{P}(\mathcal{D})$ is a constant\footnote{In this study, $\mathcal{P}(\bm{w}_h)$ is set to be  $\mathcal{N}(\bm{w}_h; \bm{0},\bm{I}_n)$.}. 
Given the learned $q_{\phi}(\bm{w}_h)$, and without additional knowledge of a human DM's unique independent decision making model, we can only postulate that the DM  follows an average model to make their decision:
\begin{equation}
   y^{h,t} =  \mathds{1}(\mathrm{E}_{q_{\phi}(\bm{w}_h)} [ \mathcal{H}(\bm{x}^t;\bm{w}_h)]>0.5)
   \label{average_prediction}
\end{equation}
Moreover, after we possess additional observations of the human DM's decision making behavior (e.g., the initial decision $y^{h,t}$ that they make), we can update our belief of the DM's independent decision making model from  the general parameter distribution $q_{\phi}(\bm{w}_h)$ in order to 
align with the observed human behavior:
\begin{equation}
\hat{q}_{\phi}(\bm{w}_h) \propto q_{\phi}(\bm{w}_h) \cdot \mathds{1}(\mathds{1}( \mathcal{H}(\bm{x}^t;\bm{w}_h) \geq 0.5 )  = y^{h,t})
\label{sample}
\end{equation}
Without loss of generality, in this study, 
we assumed that humans' decision making model $\mathcal{H}(\bm{x}^t;\bm{w}_h)$ follows the form of logistic model:
\begin{equation}
    \mathcal{H}(\bm{x}^t;\bm{w}_h) = \text{sigmoid}(\bm{w}_h \cdot \bm{x}^t)
\end{equation}

\subsection{Quantify the Nudging Effects of AI Assistance}
Inspired by a recent computational framework for understanding and predicting the effects of nudges~\cite{callaway2022optimal}, 
in this study, we introduce a computational framework 
to provide an interpretable and quantitative characterization of the influence of diverse forms of AI assistance on human decision makers, which enables us to predict human behavior in AI-assisted decision making. 
The core idea of this framework is to conceptualize the AI assistance as a ``nudge'' to the human decision making
process, 
such that it can modify how the human DM weighs different information in their decision making and alter their independent decision model accordingly. Depending on the type of AI assistance used, this alternation could be operationalized as the human DM changing their belief in the relevance of certain task feature to their decisions, or as the human DM redirecting their attention to certain task feature when making their decisions.

\begin{figure}[t]
    \centering
        \includegraphics[width=0.48\textwidth]{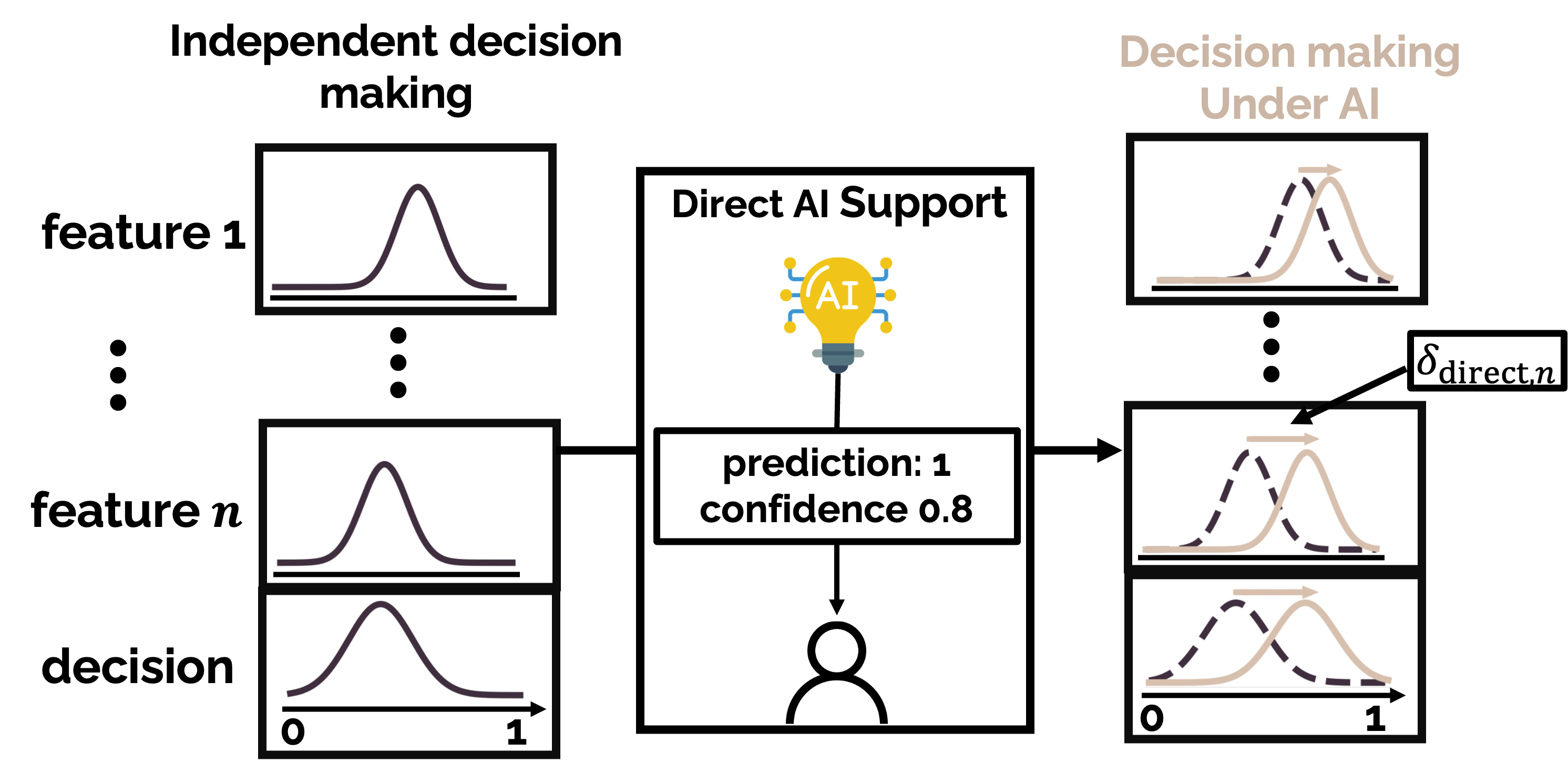}
    \caption{The illustration of how \emph{immediate assistance} nudges human decision makers.}
    \label{fig:ill_direct}
\end{figure}

\subsubsection{Immediate Assistance.}
As shown in Figure~\ref{fig:ill_direct}, in this scenario, human DMs are directly presented with the AI model's decision recommendation $y^{m,t}$ and confidence $c^{m,t}$ before they deliberate on the task trial $\bm{x}^t$. Human DMs may consciously or unconsciously incorporate the AI recommendation into their final decisions. The specific influence of AI on a human DM may largely vary with the DM's inherent attitudes towards AI.  
For example, DMs with a high tendency to trust AI may simply adopt AI's recommendation while skeptical DMs may simply adopt the opposite decision than what AI suggests. 
In less extreme cases, DMs are not simply trust or distrust the AI recommendation, 
but the AI recommendation may change their belief of the relevance/importance of different task features so that they can either align more with the AI recommendation $\hat{y}^{t}$, or deviate more from the AI recommendation  $\hat{y}^{t}$. The magnitude of this adjustment may be controlled by the AI model's confidence level $c^{m,t}$.
Therefore, given any human DM whose independent decision making model is decided by $\bm{w}_h \sim q_{\phi}(\bm{w}_h)$, with the information $y^{m,t}$ and $c^{m,t}$, the DM's final decision $\hat{y}^{t}$ would be nudged by $y^{m,t}$  and $c^{m,t}$ as:
\begin{small}
\begin{equation}
\begin{split}
      \hat{y}^{t} =
    \mathds{1}(\mathrm{E}_{q_{\phi}(\bm{w}_h)} [\mathcal{H}(\bm{x}^t;\bm{w}_h + (2y^{m,t} -1)c^{m,t}\bm{\delta}_{\text{direct}})]>0.5)
\end{split}
\end{equation}
\end{small}
\noindent where $\bm{\delta}_{\text{direct}} \in \mathcal{R}^n$ ($\delta_{\text{direct},i} \cdot \delta_{\text{direct},j} \geq 0,  \forall i, j \in \{1, \dots, n\}
$) represents the updates in the DM's belief of the relevance of different task features after receiving the immediate AI assistance. Note that $\delta_{\text{direct},i}>0$ indicates that the DM has a disposition to trust the AI (hence they update their belief of task features' relevance to align more with the AI recommendation), whereas $\delta_{\text{direct},i}<0$  suggests that the DM has a tendency to distrust AI (hence they update their belief of task features' relevance to deviate more from the AI recommendation). $c^{m,t}$ moderates the magnitude of the update, 
and $y^{m,t}$ controls the direction of the update. For example, if $y^{m,t} = 1$, $\delta_{\text{direct},i}>0$ (or $\delta_{\text{direct},i}<0$) increases (or decreases) the chance of the final decision $\hat{y}^t$ being $1$ compared to that of the DM's independent decision $y^{h, t}$. 
Conversely,
If $y^{m,t} = -1$, $\delta_{\text{direct},i}>0$ (or $\delta_{\text{direct},i}<0$) increases (or decreases) the chance of the final decision $\hat{y}^t$ being $-1$
compared to that of the DM's independent decision $y^{h, t}$.
\begin{figure}[th]
    \centering
        \includegraphics[width=0.48\textwidth]{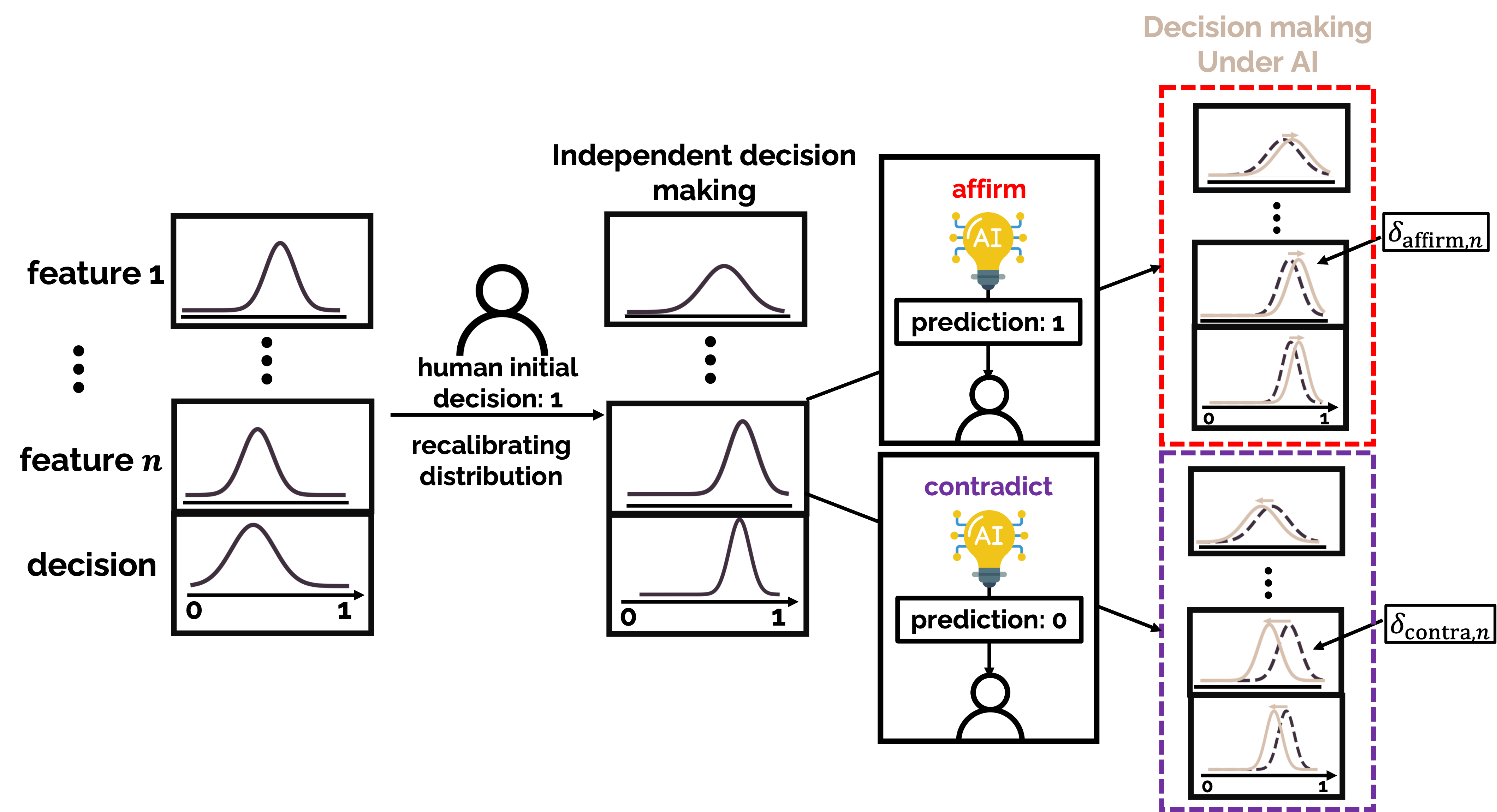}
  
    \caption{The illustration of how {\em delayed recommendation} nudges human decision makers. }
    \label{fig:ill_two_step}
\end{figure}

\subsubsection{Delayed Recommendation.}
As shown in Figure~\ref{fig:ill_two_step}, in this scenario, human DMs are required to deliberate and make their initial decision $y^{h,t}$ on task trial $\bm{x}^t$ before the AI model's decision recommendation is provided. The observed human DM's initial decision $y^{h,t}$ can be used to update our belief of the DM's independent decision making model.  
Specifically, after observing the human DM's initial decision $y^{h,t}$, an adjustment is made to the distribution of the human DM's independent decision model $q_{\phi}(\bm{w}_h)$ to filter out decision models that are inconsistent with the observed decision, yielding $\hat{q}_{\phi}(\bm{w}_h)$ as given by Eq.~\ref{sample}.
Then, as the DM compares their initial decision $y^{h,t}$ with the AI model's decision recommendation $y^{m,t}$, 
two scenarios may arise:
\begin{enumerate}
    \item \emph{AI affirms human decision} ($y^{h,t} = y^{m,t}$): For DMs who trust AI, this agreement can boost their confidence in their initial decision $y^{h,t}$. Conversely, for DMs who are skeptical of AI, they may become less confident in their own judgement due to this agreement. 
    
    \item \emph{AI contradicts human decision} ($y^{h,t} \neq y^{m,t} $): 
    DMs who tend to trust AI might reflect on their initial decision and could be inclined towards switching to $y^{m,t}$. On the other hand, DMs who are skeptical of AI may be more inclined to stand by their own judgement $y^{h,t}$.

\end{enumerate}
Depending on which scenario that the DM encounters, we model the DM's final decision $\hat{y}^{t}$ as follows: 

\begin{small}
    \begin{equation}
\begin{split}
    \hat{y}^{t} =  \mathds{1}( \mathrm{E}_{\hat{q}_{\phi}(\bm{w}_h)}  [ &\mathcal{H}(\bm{x}^t;\bm{w}_h + 
     (2y^{m,t} -1) (\mathds{1}(y^{m,t} = y^{h,t})\bm{\delta}_{\text{affirm}}+\\
     &\mathds{1}(y^{m,t} \neq y^{h,t})\bm{\delta}_{\text{contra}}))]>0.5)
\end{split}
\end{equation}
\end{small}
\noindent Here, \( \bm{\delta}_{\text{affirm}}, \bm{\delta}_{\text{contra}} \in \mathcal{R}^n \) ($ \delta_{\text{affirm},i} \cdot \delta_{\text{affirm},j} \geq 0, \; \delta_{\text{contra},i} \cdot \delta_{\text{contra},j} \geq 0, \; \forall i, j \in \{1, \dots, n\}$) represent the updates in the DM's belief of the relevance of different task features after seeing AI confirms their judgement ($y^{m,t} = y^{h,t}$) and AI contradicts their judgement ($y^{m,t} \neq y^{h,t}$), respectively.
\begin{figure}[ht]
    \centering
        \includegraphics[width=0.48\textwidth]{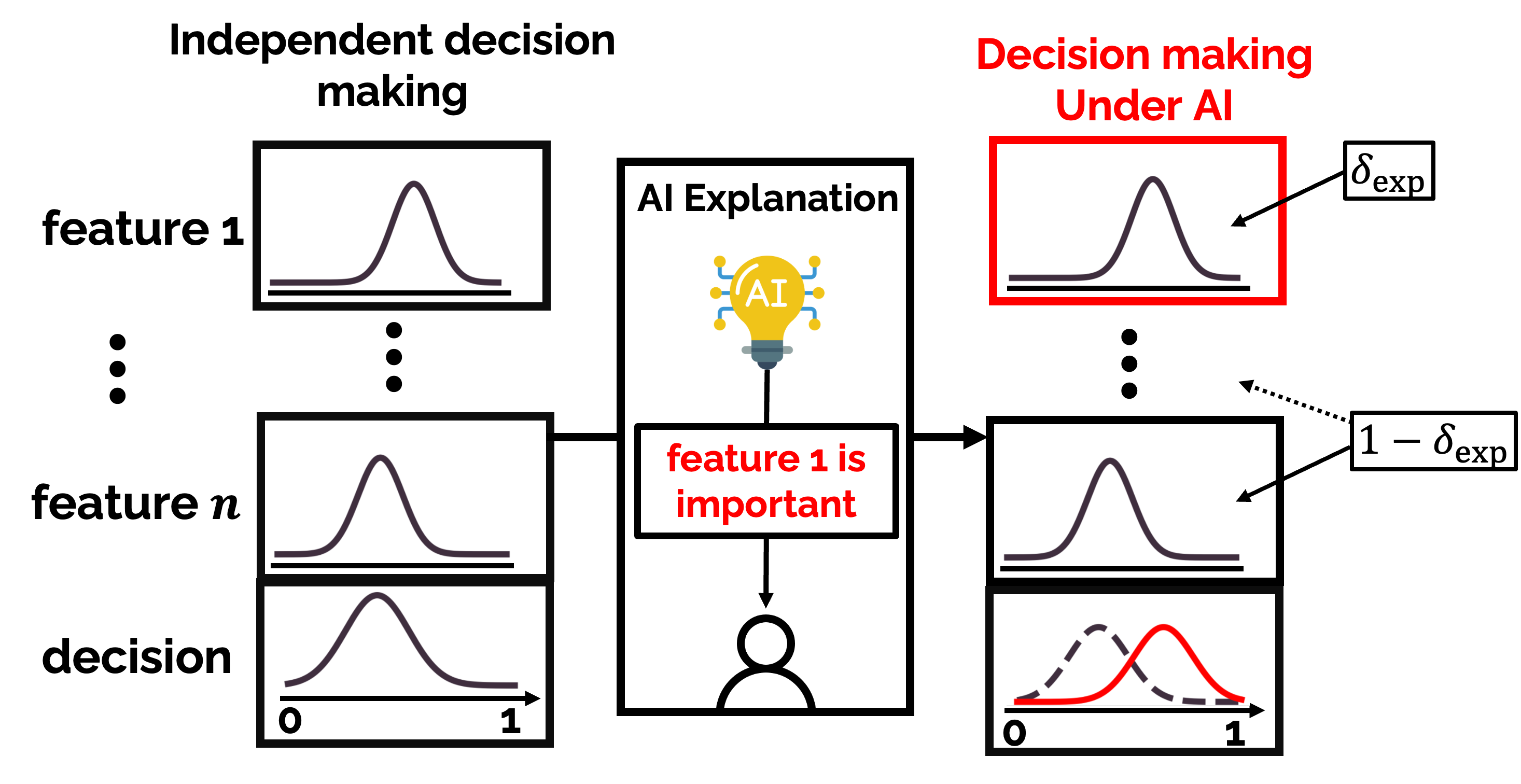}
    \caption{The illustration of how \emph{explanation only} nudges human decision makers. 
    }
    \label{fig:ill_exp}
\end{figure}

\subsubsection{Explanation Only.}
As shown in Figure~\ref{fig:ill_exp}, in this scenario, human DMs are  presented with the AI explanation $\bm{e}^{t}$, which highlights the critical features of the task trial $\bm{x}^t$. These explanations may nudge human DMs to redirect their attention to these highlighted features when forming their final decision $\hat{y}^t$. Intuitively, information that is marked as important might be prioritized by DMs, while other information may be overlooked. As such, the highlighted task features may exert a more salient influence on the DM's final decision $\hat{y}^t$:
\begin{small}
\begin{equation}
\begin{split}
    \hat{y}^{t} = \mathds{1}(\mathrm{E}_{q_{\phi}(\bm{w}_h)} & [\delta_{\text{exp}} \mathcal{H}(\bm{e}^t \odot \boldsymbol{x}^t;\bm{w}_h) + \\
    &(1-\delta_{\text{exp}})\mathcal{H}((\bm{1}-\bm{e}^t) \odot \boldsymbol{x}^t;\bm{w}_h)]>0.5)
\end{split}
\end{equation}
\end{small}
\noindent where $\odot$ is the element-wise product. $\delta_{\text{exp}} \in [0,1]$ quantifies the degree to which the human DMs redirect their attention to the highlighted features after seeing the AI explanation, 
with larger values 
indicating that the DM puts a greater emphasis on the highlighted information.

\section{Human-Subject Experiment}
To evaluate the proposed framework, 
we conducted a human-subject experiment to collect behavior data from real human decision makers in AI-assisted decision making under different forms of AI assistance.

\vspace{2pt}
\noindent \textbf{Decision Making Tasks.} The decision making task we used in our experiment was to predict diabetes in patients based on demographic information and medical history. Specifically, in each task, the subject is presented with a patient profile encompassing six features: gender, age, history of heart disease, Body Mass Index (BMI), HbA1c level, and blood glucose level at a given time interval. The subject was then asked to determine whether this patient has diabetes. The patient profiles were randomly sampled from the diabetes prediction dataset~\cite{mustafaz_dia}. 

\vspace{2pt}
\noindent \textbf{Experimental Treatments.} We created four treatments in this experiment. One of these is an \emph{independent} treatment where human subjects complete decision making tasks without any AI assistance.  In the other three treatments, human subjects receive one of the three forms of AI assistance as what we introduced earlier. The AI assistant used in the experiment was based on a boosting tree model trained on the diabetes prediction dataset. The accuracy of the AI model was 87\%.  In the \emph{Explanation only} treatment,  we used SHAP~\cite{lundberg2017unified} to explain the predictions of the boosting tree model. The two most influential features are highlighted as the AI model's explanation.

\vspace{2pt}
\noindent\textbf{Experimental Procedure.} We posted our experiment on Amazon Mechanical Turk (MTurk) as a human intelligence task (HIT) and recruited MTurk workers as our subjects. Upon arrival, we randomly assigned each subject to one of the four treatments. 
Subjects started the HIT by completing a tutorial that described the diabetes prediction task that they needed to work on in the HIT and the meaning of each feature they would see in a patient’s profile. To familiarize subjects with the task, we initially asked them to complete five training tasks. During these training tasks, subjects made diabetes predictions without AI assistance, and we immediately provided them with the correct answers and the end of each task.
The real experiment began after the subject completed the training tasks. Specifically,  subjects were asked to complete a total of 30 tasks, which were randomly sampled from a pool of 500 task instances. After subjects completed all 30 tasks, subjects were asked to undertake a 3-item Cognitive Reflection Test (CRT)~\cite{Frederick2005CognitiveRA}, intended to assess the subject's tendency in engaging with intuitve vs. reflective thinking. 
We offered a base payment of \$1.2 for the HIT. The HIT was open to US-based workers only, and each worker can complete the HIT once. We further included an attention check question within the HIT, where subjects were required to select a randomly determined option. Data collected from subjects who successfully passed the attention check were considered valid for our study (see the supplementary materials for more details of the human-subject experiment).

\section{Evaluations}
\begin{table*}
\centering
\small
\begin{tblr}{
  cells = {c},
  cell{1}{1} = {r=2}{},
  cell{1}{2} = {c=3}{},
  cell{1}{5} = {c=3}{},
  cell{1}{8} = {c=3}{},
  vline{2} = {1-8}{},
  vline{3,6} = {1}{dashed},
  vline{3-4,6-7,9-10} = {2-8}{dotted},
  vline{5,8} = {1-8}{dashed},
  hline{1,3,9} = {-}{},
  hline{2} = {2-10}{},
}
Treatment           & Immediate assistance&          &       & Delayed recommendation &          &       & Explanation only &          &       \\
                    & NLL $\downarrow$                        & Accuracy $\uparrow$ & F1 $\uparrow$    & NLL $\downarrow$               & Accuracy $\uparrow$ & F1 $\uparrow$    & NLL $\downarrow$            & Accuracy $\uparrow$ & F1 $\uparrow$    \\
Logistic Regression & 0.522                      & 0.753    & 0.789 & 0.446             & 0.809    & 0.782 & \textbf{0.549}          & \textbf{0.728}   & 0.767\\
XGBoost             & 0.533                           & 0.768         & 0.737      & 0.472             & 0.812    & 0.797 & 0.617          & 0.711    & 0.753 \\
MLP                 & 0.656                           & 0.753        & 0.729      & 0.554             & 0.777    & 0.751 & 0.606          & 0.686    & 0.778 \\
SVM                 & 0.530                           & 0.754          & 0.707      & 0.461             & 0.791    & 0.758 & 0.603          & 0.721   & 0.743 \\
Utility                 & 0.573                          & 0.739          & 0.779      & -             & -    & - & -         & -   & - \\
Ours                & \textbf{0.435}                          & \textbf{0.800}         & \textbf{0.818}      & \textbf{0.413}             & \textbf{0.825}    & \textbf{0.812} & 0.563          & 0.715    & \textbf{0.791} 
\end{tblr}
\caption{Comparing the performance of proposed method with baseline methods on three forms of AI assistance, in terms of NLL, Accuracy, and F1-score. ``$\downarrow$'' denotes the lower
the better, ``$\uparrow$'' denotes the higher the better. Best result in each column is
highlighted in bold. All results are averaged over 5 runs. ``-'' means the method can not be applied in this scenario. 
}
\label{tab:performance1}
\end{table*}

After filtering the inattentive subjects, we obtained valid data from 202 subjects in our experiment (\emph{Independent}: 53, \emph{Immediate assistance}: 50, \emph{Delayed recommendation}: 53, \emph{Explanation only}: 46). Below, we conduct our evaluation using the behavior data collected from these valid subjects.
\subsection{Model Training and Baselines}
We first learned the general parameter distribution $q_{\phi}(\bm{w}_h)$ of human DMs' independent decision making model
utilizing the data collected in the \emph{Independent} treatment. Through 5-fold cross-validation, we found the average accuracy in predicting an average human DM's independent decision
using $q_{\phi}(\bm{w}_h)$ and Eq.~\ref{average_prediction}  is $0.673$. In the following evaluations, we used $q_{\phi}(\bm{w}_h)$ to reflect the human DM's independent decision making model, and used the maximum likelihood estimation to learn how different forms of AI assistance nudge each individual human DM to modify their decision making model.

Specifically, to evaluate the performance of the proposed framework, for each human subject, we randomly split the behavior data collected from them into training 
(50\%) and test (50\%) sets. We computed the average negative log-likelihood (NLL) to measure how well different models capture the likelihood of subjects making their final decisions under the influence of AI assistance,  
and we averaged the NLL values across all subjects in each treatment. A lower mean NLL indicates a better prediction performance. In addition, we also employed F1-score and Accuracy scores to evaluate the performance of different models. For both these metrics, higher scores denote better performance. To ensure the robustness of evaluations, all experiments were repeated 5 times, and the average performance across these repetitions was reported.

We consider utility-based  model proposed by~\citeauthor{wang2022will}~\shortcite{wang2022will} and  a few standard supervised learning models as baselines in evaluations, including Logistic Regression, XGBoost, Multi-Layer Perceptron (MLP), and Support Vector Machines (SVM). These supervised learning models directly predicts human DMs' final decisions $\hat{y}^{h,t}$ in a decision task based on various features:
\begin{enumerate}
    \item \emph{Immediate assistance:} task trial features $\bm{x}^t$, as well as the AI model's decision recommendation $y^{m,t}$ and confidence $c^{m,t}$ in the task trial.
    \item  \emph{Delayed recommendation:} task trial features $\bm{x}^t$, human DMs' initial decision $y^{h,t}$, 
    and the AI model's decision recommendation $y^{m,t}$ in the task trial.

    \item \emph{Explanation only:} task trial features $\bm{x}^t$ and the AI explanation $\bm{e}^t$ in the task trial.
\end{enumerate}

\subsection{Comparing Model Performance}
Table~\ref{tab:performance1} presents the comparative performance of various models in predicting human DMs' decisions across three forms of AI assistance. Overall, our proposed method consistently outperforms the baseline methods in the \emph{Immediate assistance} and \emph{Delayed recommendation} scenarios across all metrics by a significant margin. For instance, within the \emph{Immediate assistance} scenario, the NLL for our method stands at a mere $0.435$, whereas the best baseline achieves an NLL of $0.522$. In the \emph{Explanation only} scenario, the performance of our method is comparable with the best-performing baseline model, logistic regression, in terms of NLL and Accuracy, and outperforms it on the F1-score. 

\begin{figure}[t]
    \centering
    \subfloat[Negative Log Likelihood]{%
        \includegraphics[width=0.235\textwidth]{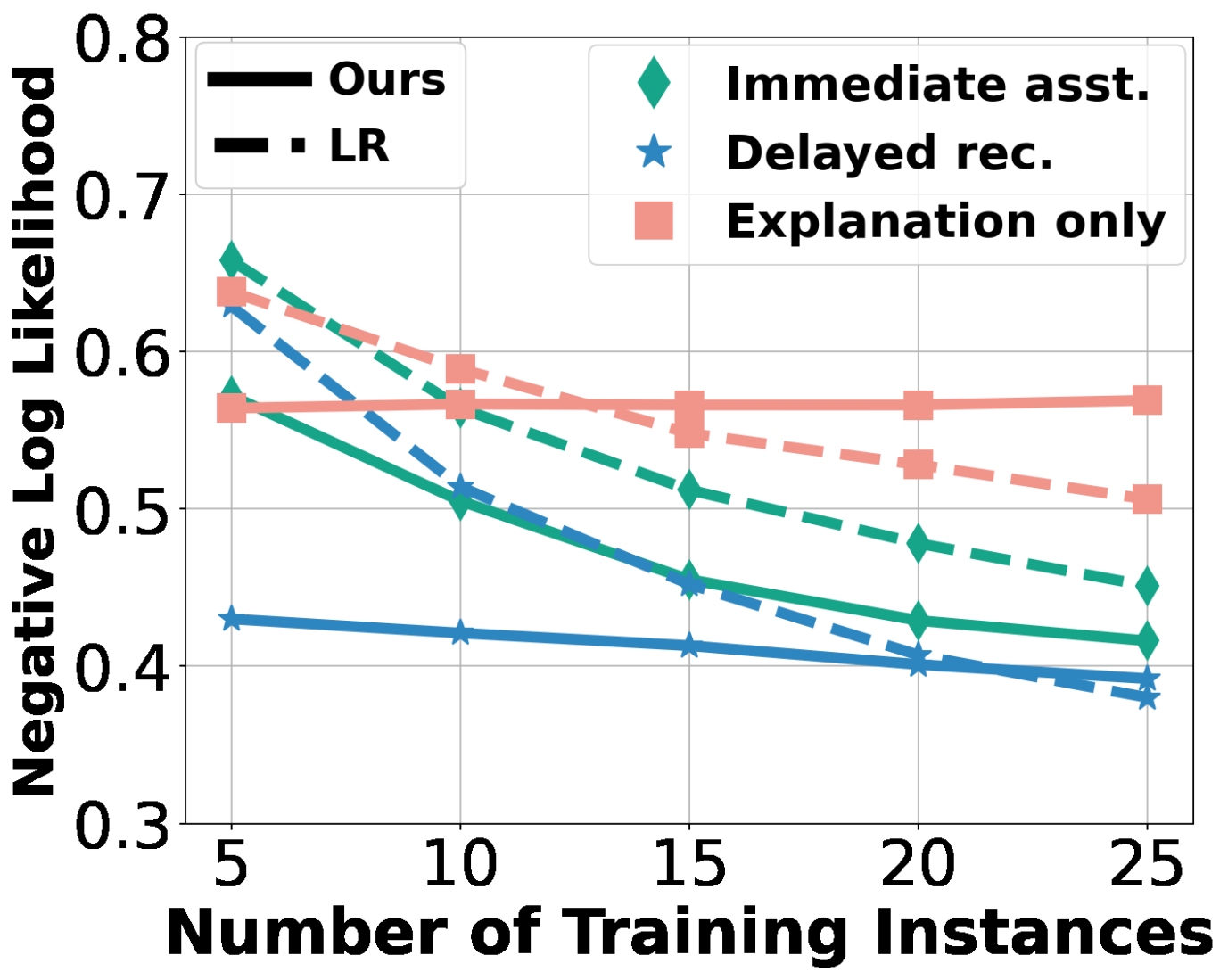}}
    \hfill
    \subfloat[F1-score]{%
        \includegraphics[width=0.235\textwidth]{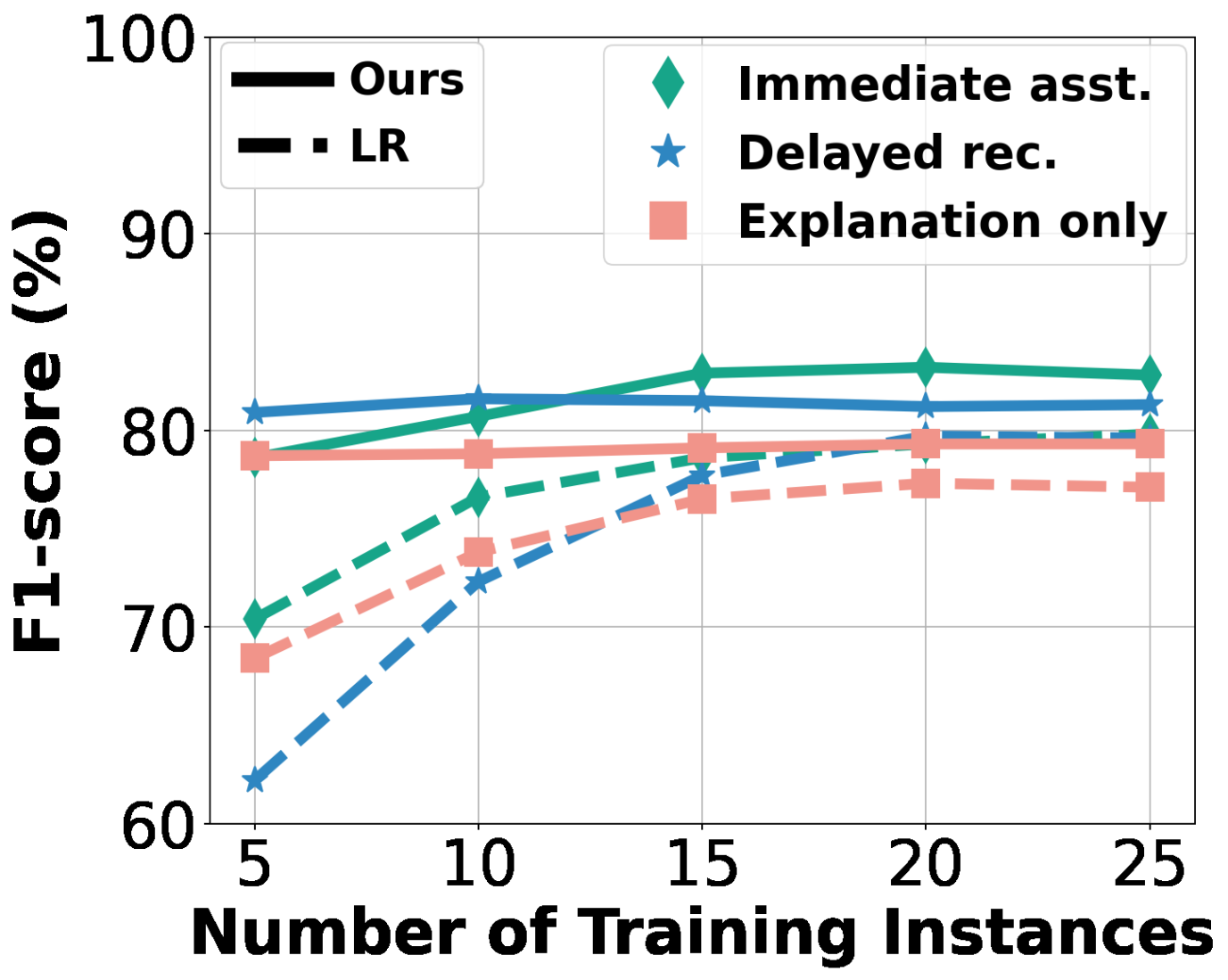}}
    \caption{Comparing the performance of our method with logistic regression models when changing the size of training data under three forms of AI assistance. }
    \label{fig:performance2}
\end{figure}

To assess the robustness of our approach, we varied the proportions of training and testing data and observed how the performance of our method changes with the training data size. 
Given the high performance of the logistic regression model,  
we selected it as the baseline model in this evaluation. As shown in Figure~\ref{fig:performance2}, our approach demonstrates consistently superior performance compared to logistic regression models across three AI-assisted decision making scenarios particularly when the number of training instances is limited. 
Specifically, the performance of our model remains robust with respect to variations in the amount of training data; it shows only a slight decrease in performance as the number of training instances reduces. In contrast, logistic regression models are highly sensitive to the size of the training data. As the number of training samples decreases, their performance degrades significantly. In other words, unlike the standard supervised learning models like logistic regression---which requires retraining from scratch for each individual human DM---our approach is endowed with the knowledge of how human DMs in general make decisions. 
This knowledge makes it possible for us to only tune the parameters of AI's nudging effects $\bm{\delta}$ on each individual human DM with a few training instances, yet still achieving comparable or even higher performance compared to the supervised learning models. 

\begin{figure*}[t]
    \centering
    \subfloat[Immediate assistance: $||\bm{\delta}_{\text{direct}}||$]{%
        \includegraphics[width=0.235\textwidth]{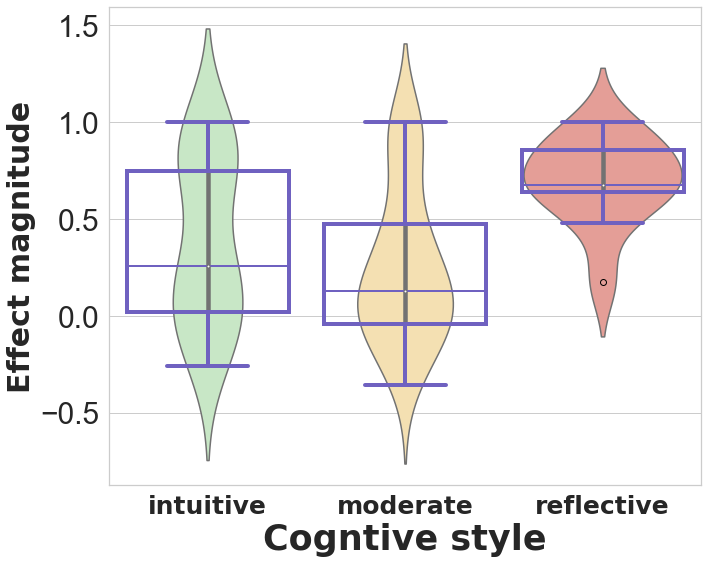}\label{fig:direct_delta}}
    \hfill
     \subfloat[Delayed recommendation ($y^{h} = y^{m}$): $||\bm{\delta}_{\text{affirm}}||$]{%
        \includegraphics[width=0.235\textwidth]{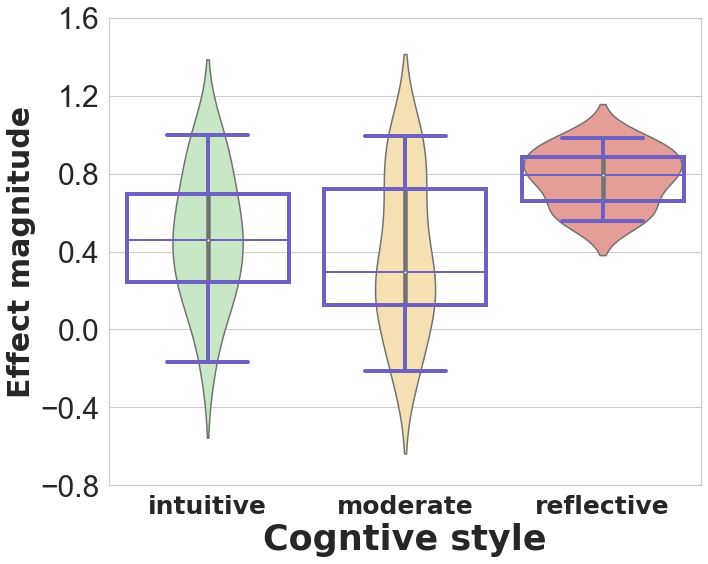}\label{fig:two_affirm_delta}}
    \hfill
     \subfloat[Delayed recommendation ($y^{h} \neq y^{m}$): $||\bm{\delta}_{\text{contra}}||$]{%
        \includegraphics[width=0.235\textwidth]{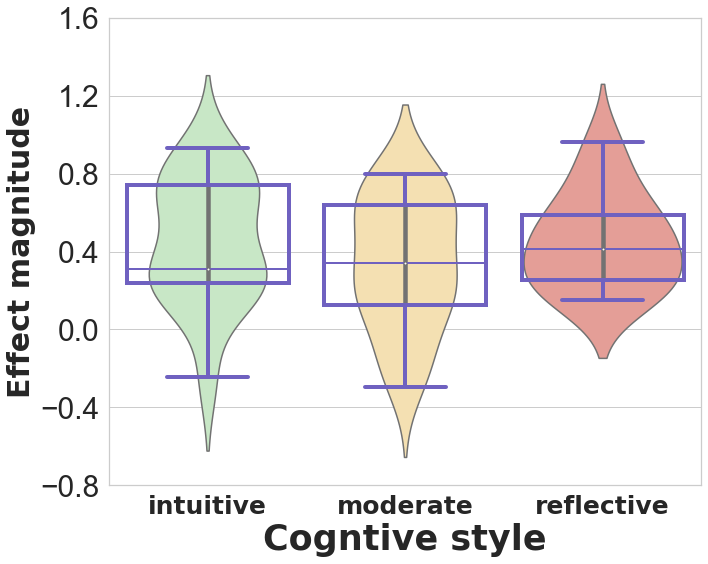}\label{fig:two_contra_delta}}
    \hfill
    \subfloat[Explanation only: $\delta_{\text{exp}}$]{%
        \includegraphics[width=0.235\textwidth]{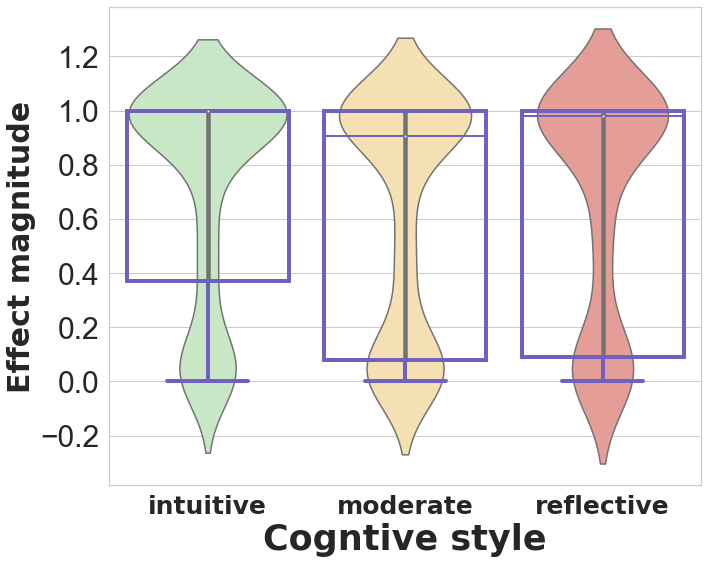}\label{fig:exp_delta}}
    \caption{Comparing the nudging effects of AI assistance on decision makers with different cognitive styles across three forms of AI assistance.}
    \label{fig:effect_comparsion}
\end{figure*}

\subsection{Examining the Nudging Effect of AI Assistance across Individuals}

We now examine how may the AI assistance nudges decision makers with different cognitive styles similarly or differently.  To do so, we compared the size of the learned nudging effects on decision makers for three forms of AI assistance. Specifically, we first used all behavior data collected from a human DM to learn the nudging effect of AI assistance on them (i.e., $\bm{\delta}$). 
We then used $\text{sign}(\bm{\delta})||\bm{\delta}||$ to represent the direction and magnitude of the nudging effects of AI assistance on the human DM ($\text{sign}(\bm{\delta})=1$ when $\forall i, \delta_i>0$; otherwise $\text{sign}(\bm{\delta})=-1$). 
To categorize the cognitive style of each human DM (i.e., each subject), we utilized 
their scores in the 3-item Cognitive Reflection Test (CRT) in the experiment---Following previous research~\cite{Frederick2005CognitiveRA}, subjects scoring 3 were classified as having a reflective thinking style, those with a score of 0 were categorized as having an intuitive thinking style, and those scoring 1 or 2 were categorized as the moderate reflective style.


Figure~\ref{fig:effect_comparsion} shows the comparison results of the nudging effects of AI assistance on DMs with different cognitive styles. To examine whether the nudging effects of AI assistance are different across DMs with different cognitive styles,  
we first used a one-way ANOVA test\footnote{Analysis of Variance (ANOVA) is a statistical test for identifying significant differences between group means.} to determine if there are statistically significant differences in the values of $\text{sign}(\bm{\delta})||\bm{\delta}||$ across different groups of DMs. If significant differences are detected, we proceed with post-hoc pairwise comparisons using Tukey's HSD test\footnote{Tukey's HSD (Honestly Significant Difference) is a post-hoc test used to determine specific differences between pairs of group means after a one-way ANOVA test has found significance.}.
Overall, our findings suggest that under the \emph{Immediate assistance} and the \emph{Delayed recommendation} scenarios (when AI affirms human decision), AI assistance exerts a larger influence on DMs with a reflective thinking style than intuitive DMs ($p<0.05$) and DMs with moderate reflective styles ($p<0.05$). One potential explanation is that reflective DMs are inclined to deliberate more extensively on tasks and 
the AI model's recommendations. 
Thus, through their interactions with the AI model, reflective DMs may have sensed the high performance of the AI model (its accuracy is 87\% for the task), making them more willing to align their decisions with the AI recommendation, especially when the AI recommendation affirms their own independent judgement. However, when the human DM's initial decision differs from the AI recommendation in the \emph{Delayed recommendation} scenario, there isn't a statistical difference in the AI's nudging effects across the three types of decision makers. In fact, by comparing the AI's nudging effects on different groups of DMs under the two conditions of the \emph{Delayed recommendation} scenario---AI affirms human decisions or contradicts human decisions---we find that reflective DMs are significantly more likely to align their final decisions with the AI recommendation when AI affirms rather than contradicting their initial judgement ($p<0.05$). In contrast, the intuitive and moderately reflective DMs do not appear to be impacted by AI significantly differently under these two conditions.

Finally, under the \emph{Explanation only} scenario, we also observe that intuitive DMs tend to place more emphasis on the features highlighted by the AI explanations. The nudging effect of AI explanations on intuitive DMs is found to be significantly greater than that on moderately reflective DMs ($p<0.05$). While the nudging effect also appears to be slightly larger for intuitive DMs compared to reflective DMs, the difference is not statistically significant.

\section{Conclusion}
In this paper, we propose a  computational framework to characterize how various AI assistance nudges humans in AI-assisted decision making. We evaluate the proposed model's performance in fitting the real human behavior data collected from a randomized experiment. Our results show that the proposed model consistently outperforms
other baselines in accurately predicting humans decisions under diverse forms of AI assistance.  Additionally, further analyses based on the proposed framework provided insights into how individuals with varying cognitive styles are impacted by AI assistance differently.

There are a few limitations of this study. 
For example, the behavior data is collected from laypeople on the diabetes
prediction task, which contains a relatively small number of features. It remains to be investigated whether the proposed model can perform well on tasks that involve many more features and thus more complex. Additionally, the AI-assisted decision scenario we examined in this study lacks sequential or temporal feedback regarding AI performance. Further exploration is required to generalize the propose framework to the sequential settings. Lastly, we assumed that the independent human decision model follows the form of logistic regression. Additional research is needed to explore how to adapt the current ways of altering humans' decision models for reflecting the nudging effect of AI assistance on human DMs to other forms of decision models.

\newpage

\section*{Acknowledgements}
We thank the support of the National Science Foundation
under grant IIS-2229876 on this work.
Any opinions, findings, conclusions, or recommendations expressed here are those of the authors alone.

\bibliography{aaai24}

\begin{thebibliography}{71}
\providecommand{\natexlab}[1]{#1}

\bibitem[{Abdul et~al.(2020)Abdul, von~der Weth, Kankanhalli, and
  Lim}]{Abdul2020COGAMMA}
Abdul, A.; von~der Weth, C.; Kankanhalli, M.~S.; and Lim, B.~Y. 2020.
\newblock COGAM: Measuring and Moderating Cognitive Load in Machine Learning
  Model Explanations.
\newblock \emph{Proceedings of the 2020 CHI Conference on Human Factors in
  Computing Systems}.

\bibitem[{Ajenaghughrure et~al.(2019)Ajenaghughrure, Sousa, Kosunen, and
  Lamas}]{ajenaghughrure2019predictive}
Ajenaghughrure, I.~B.; Sousa, S.~C.; Kosunen, I.~J.; and Lamas, D. 2019.
\newblock Predictive model to assess user trust: a psycho-physiological
  approach.
\newblock In \emph{Proceedings of the 10th Indian conference on human-computer
  interaction}, 1--10.

\bibitem[{Alqaraawi et~al.(2020)Alqaraawi, Schuessler, Wei{\ss}, Costanza, and
  Bianchi-Berthouze}]{Alqaraawi2020EvaluatingSM}
Alqaraawi, A.; Schuessler, M.; Wei{\ss}, P.; Costanza, E.; and
  Bianchi-Berthouze, N. 2020.
\newblock Evaluating saliency map explanations for convolutional neural
  networks: a user study.
\newblock \emph{Proceedings of the 25th International Conference on Intelligent
  User Interfaces}.

\bibitem[{Anik and Bunt(2021)}]{Anik2021DataCentricEE}
Anik, A.~I.; and Bunt, A. 2021.
\newblock Data-Centric Explanations: Explaining Training Data of Machine
  Learning Systems to Promote Transparency.
\newblock \emph{Proceedings of the 2021 CHI Conference on Human Factors in
  Computing Systems}.

\bibitem[{Bansal et~al.(2021)Bansal, Nushi, Kamar, Horvitz, and
  Weld}]{Bansal2021IsTM}
Bansal, G.; Nushi, B.; Kamar, E.; Horvitz, E.; and Weld, D.~S. 2021.
\newblock Is the Most Accurate AI the Best Teammate? Optimizing AI for
  Teamwork.
\newblock In \emph{AAAI Conference on Artificial Intelligence}.

\bibitem[{Bansal et~al.(2020)Bansal, Wu, Zhou, Fok, Nushi, Kamar, Ribeiro, and
  Weld}]{Bansal2020DoesTW}
Bansal, G.; Wu, T.~S.; Zhou, J.; Fok, R.; Nushi, B.; Kamar, E.; Ribeiro, M.~T.;
  and Weld, D.~S. 2020.
\newblock Does the Whole Exceed its Parts? The Effect of AI Explanations on
  Complementary Team Performance.
\newblock \emph{Proceedings of the 2021 CHI Conference on Human Factors in
  Computing Systems}.

\bibitem[{Brown et~al.(2019)Brown, Chouldechova, Putnam-Hornstein, Tobin, and
  Vaithianathan}]{Brown2019TowardAA}
Brown, A.; Chouldechova, A.; Putnam-Hornstein, E.; Tobin, A.; and
  Vaithianathan, R. 2019.
\newblock Toward Algorithmic Accountability in Public Services: A Qualitative
  Study of Affected Community Perspectives on Algorithmic Decision-making in
  Child Welfare Services.
\newblock \emph{Proceedings of the 2019 CHI Conference on Human Factors in
  Computing Systems}.

\bibitem[{Buccinca et~al.(2020)Buccinca, Lin, Gajos, and
  Glassman}]{Buccinca2020ProxyTA}
Buccinca, Z.; Lin, P.; Gajos, K.~Z.; and Glassman, E.~L. 2020.
\newblock Proxy tasks and subjective measures can be misleading in evaluating
  explainable AI systems.
\newblock \emph{Proceedings of the 25th International Conference on Intelligent
  User Interfaces}.

\bibitem[{Bu{\c{c}}inca, Malaya, and Gajos(2021)}]{Buccinca2021ToTO}
Bu{\c{c}}inca, Z.; Malaya, M.~B.; and Gajos, K.~Z. 2021.
\newblock To trust or to think: cognitive forcing functions can reduce
  overreliance on AI in AI-assisted decision-making.
\newblock \emph{Proceedings of the ACM on Human-Computer Interaction},
  5(CSCW1): 1--21.

\bibitem[{Cai, Jongejan, and Holbrook(2019)}]{Cai2019TheEO}
Cai, C.~J.; Jongejan, J.; and Holbrook, J. 2019.
\newblock The effects of example-based explanations in a machine learning
  interface.
\newblock \emph{Proceedings of the 24th International Conference on Intelligent
  User Interfaces}.

\bibitem[{Cai et~al.(2019{\natexlab{a}})Cai, Reif, Hegde, Hipp, Kim, Smilkov,
  Wattenberg, Viegas, Corrado, Stumpe, and Terry}]{10.1145/3290605.3300234}
Cai, C.~J.; Reif, E.; Hegde, N.; Hipp, J.; Kim, B.; Smilkov, D.; Wattenberg,
  M.; Viegas, F.; Corrado, G.~S.; Stumpe, M.~C.; and Terry, M.
  2019{\natexlab{a}}.
\newblock Human-Centered Tools for Coping with Imperfect Algorithms During
  Medical Decision-Making.
\newblock In \emph{Proceedings of the 2019 CHI Conference on Human Factors in
  Computing Systems}, CHI '19, 1–14. New York, NY, USA: Association for
  Computing Machinery.
\newblock ISBN 9781450359702.

\bibitem[{Cai et~al.(2019{\natexlab{b}})Cai, Reif, Hegde, Hipp, Kim, Smilkov,
  Wattenberg, Vi{\'e}gas, Corrado, Stumpe, and Terry}]{Cai2019HumanCenteredTF}
Cai, C.~J.; Reif, E.; Hegde, N.; Hipp, J.~D.; Kim, B.; Smilkov, D.; Wattenberg,
  M.; Vi{\'e}gas, F.~B.; Corrado, G.~S.; Stumpe, M.~C.; and Terry, M.
  2019{\natexlab{b}}.
\newblock Human-Centered Tools for Coping with Imperfect Algorithms During
  Medical Decision-Making.
\newblock \emph{Proceedings of the 2019 CHI Conference on Human Factors in
  Computing Systems}.

\bibitem[{Callaway, Hardy, and Griffiths(2022)}]{callaway2022optimal}
Callaway, F.; Hardy, M.; and Griffiths, T. 2022.
\newblock Optimal nudging for cognitively bounded agents: A framework for
  modeling, predicting, and controlling the effects of choice architectures.

\bibitem[{Cheng et~al.(2019)Cheng, Wang, Zhang, O'Connell, Gray, Harper, and
  Zhu}]{Cheng2019ExplainingDA}
Cheng, H.~F.; Wang, R.; Zhang, Z.; O'Connell, F.; Gray, T.; Harper, F.~M.; and
  Zhu, H. 2019.
\newblock Explaining Decision-Making Algorithms through UI: Strategies to Help
  Non-Expert Stakeholders.
\newblock \emph{Proceedings of the 2019 CHI Conference on Human Factors in
  Computing Systems}.

\bibitem[{Chromik et~al.(2021)Chromik, Eiband, Buchner, Kr{\"u}ger, and
  Butz}]{Chromik2021ITI}
Chromik, M.; Eiband, M.; Buchner, F.; Kr{\"u}ger, A.; and Butz, A.~M. 2021.
\newblock I Think I Get Your Point, AI! The Illusion of Explanatory Depth in
  Explainable AI.
\newblock \emph{26th International Conference on Intelligent User Interfaces}.

\bibitem[{Das and Chernova(2020)}]{Das2020LeveragingRT}
Das, D.; and Chernova, S. 2020.
\newblock Leveraging rationales to improve human task performance.
\newblock \emph{Proceedings of the 25th International Conference on Intelligent
  User Interfaces}.

\bibitem[{De-Arteaga, Fogliato, and Chouldechova(2020)}]{DeArteaga2020ACF}
De-Arteaga, M.; Fogliato, R.; and Chouldechova, A. 2020.
\newblock A Case for Humans-in-the-Loop: Decisions in the Presence of Erroneous
  Algorithmic Scores.
\newblock \emph{Proceedings of the 2020 CHI Conference on Human Factors in
  Computing Systems}.

\bibitem[{Desmond et~al.(2021)Desmond, Ashktorab, Brachman, Brimijoin,
  Duesterwald, Dugan, Finegan-Dollak, Muller, Joshi, Pan, and
  Sharma}]{Desmond2021IncreasingTS}
Desmond, M.; Ashktorab, Z.; Brachman, M.; Brimijoin, K.; Duesterwald, E.;
  Dugan, C.; Finegan-Dollak, C.; Muller, M.~J.; Joshi, N.~N.; Pan, Q.; and
  Sharma, A. 2021.
\newblock Increasing the Speed and Accuracy of Data Labeling Through an AI
  Assisted Interface.
\newblock \emph{26th International Conference on Intelligent User Interfaces}.

\bibitem[{Dodge et~al.(2019)Dodge, Liao, Zhang, Bellamy, and
  Dugan}]{Dodge2019ExplainingMA}
Dodge, J.; Liao, Q.~V.; Zhang, Y.; Bellamy, R. K.~E.; and Dugan, C. 2019.
\newblock Explaining models: an empirical study of how explanations impact
  fairness judgment.
\newblock \emph{Proceedings of the 24th International Conference on Intelligent
  User Interfaces}.

\bibitem[{Feng and Boyd-Graber(2018)}]{Feng2018WhatCA}
Feng, S.; and Boyd-Graber, J.~L. 2018.
\newblock What can AI do for me?: evaluating machine learning interpretations
  in cooperative play.
\newblock \emph{Proceedings of the 24th International Conference on Intelligent
  User Interfaces}.

\bibitem[{Fogliato et~al.(2022)Fogliato, Chappidi, Lungren, Fitzke, Parkinson,
  Wilson, Fisher, Horvitz, Inkpen, and Nushi}]{Fogliato2022WhoGF}
Fogliato, R.; Chappidi, S.; Lungren, M.~P.; Fitzke, M.; Parkinson, M.; Wilson,
  D.~U.; Fisher, P.; Horvitz, E.; Inkpen, K.; and Nushi, B. 2022.
\newblock Who Goes First? Influences of Human-AI Workflow on Decision Making in
  Clinical Imaging.
\newblock \emph{Proceedings of the 2022 ACM Conference on Fairness,
  Accountability, and Transparency}.

\bibitem[{Frederick(2005)}]{Frederick2005CognitiveRA}
Frederick, S. 2005.
\newblock Cognitive Reflection and Decision Making.
\newblock \emph{Journal of Economic Perspectives}, 19: 25--42.

\bibitem[{Gajos and Mamykina(2022)}]{Gajos2022DoPE}
Gajos, K.~Z.; and Mamykina, L. 2022.
\newblock Do People Engage Cognitively with AI? Impact of AI Assistance on
  Incidental Learning.
\newblock \emph{27th International Conference on Intelligent User Interfaces}.

\bibitem[{Gomez et~al.(2020)Gomez, Holter, Yuan, and Bertini}]{Gomez2020ViCEVC}
Gomez, O.; Holter, S.; Yuan, J.; and Bertini, E. 2020.
\newblock ViCE: visual counterfactual explanations for machine learning models.
\newblock \emph{Proceedings of the 25th International Conference on Intelligent
  User Interfaces}.

\bibitem[{Green and Chen(2019{\natexlab{a}})}]{Green2019DisparateIA}
Green, B.; and Chen, Y. 2019{\natexlab{a}}.
\newblock Disparate Interactions: An Algorithm-in-the-Loop Analysis of Fairness
  in Risk Assessments.
\newblock \emph{Proceedings of the Conference on Fairness, Accountability, and
  Transparency}.

\bibitem[{Green and Chen(2019{\natexlab{b}})}]{Green2019ThePA}
Green, B.; and Chen, Y. 2019{\natexlab{b}}.
\newblock The Principles and Limits of Algorithm-in-the-Loop Decision Making.
\newblock \emph{Proceedings of the ACM on Human-Computer Interaction}, 3: 1 --
  24.

\bibitem[{Grgic-Hlaca, Engel, and Gummadi(2019)}]{GrgicHlaca2019HumanDM}
Grgic-Hlaca, N.; Engel, C.; and Gummadi, K.~P. 2019.
\newblock Human Decision Making with Machine Assistance: An Experiment on
  Bailing and Jailing.
\newblock \emph{DecisionSciRN: Decision-Making in the Legal Field (Topic)}.

\bibitem[{Grgi\'{c}-Hla\v{c}a, Engel, and Gummadi(2019)}]{10.1145/3359280}
Grgi\'{c}-Hla\v{c}a, N.; Engel, C.; and Gummadi, K.~P. 2019.
\newblock Human Decision Making with Machine Assistance: An Experiment on
  Bailing and Jailing.
\newblock \emph{Proc. ACM Hum.-Comput. Interact.}, 3(CSCW).

\bibitem[{Guo et~al.(2019)Guo, Du, Malik, Koh, Kim, Liu, Kim, Zha, and
  Cao}]{Guo2019VisualizingUA}
Guo, S.; Du, F.; Malik, S.; Koh, E.; Kim, S.; Liu, Z.; Kim, D.; Zha, H.; and
  Cao, N. 2019.
\newblock Visualizing Uncertainty and Alternatives in Event Sequence
  Predictions.
\newblock \emph{Proceedings of the 2019 CHI Conference on Human Factors in
  Computing Systems}.

\bibitem[{Harrison et~al.(2020)Harrison, Hanson, Jacinto, Ramirez, and
  Ur}]{Harrison2020AnES}
Harrison, G.; Hanson, J.; Jacinto, C.; Ramirez, J.; and Ur, B. 2020.
\newblock An empirical study on the perceived fairness of realistic, imperfect
  machine learning models.
\newblock \emph{Proceedings of the 2020 Conference on Fairness, Accountability,
  and Transparency}.

\bibitem[{Jacobs et~al.(2021)Jacobs, He, Pradier, Lam, Ahn, McCoy, Perlis,
  Doshi-Velez, and Gajos}]{Jacobs2021DesigningAF}
Jacobs, M.~L.; He, J.; Pradier, M.~F.; Lam, B.; Ahn, A.~C.; McCoy, T.~H.;
  Perlis, R.~H.; Doshi-Velez, F.; and Gajos, K.~Z. 2021.
\newblock Designing AI for Trust and Collaboration in Time-Constrained Medical
  Decisions: A Sociotechnical Lens.
\newblock \emph{Proceedings of the 2021 CHI Conference on Human Factors in
  Computing Systems}.

\bibitem[{Kocielnik, Amershi, and Bennett(2019)}]{Kocielnik2019WillYA}
Kocielnik, R.; Amershi, S.; and Bennett, P.~N. 2019.
\newblock Will You Accept an Imperfect AI?: Exploring Designs for Adjusting
  End-user Expectations of AI Systems.
\newblock \emph{Proceedings of the 2019 CHI Conference on Human Factors in
  Computing Systems}.

\bibitem[{Kumar et~al.(2021)Kumar, Patel, Benjamin, and
  Steyvers}]{kumar2021explaining}
Kumar, A.; Patel, T.; Benjamin, A.~S.; and Steyvers, M. 2021.
\newblock Explaining algorithm aversion with metacognitive bandits.
\newblock In \emph{Proceedings of the annual meeting of the cognitive science
  society}, volume~43.

\bibitem[{Kunkel et~al.(2019)Kunkel, Donkers, Michael, Barbu, and
  Ziegler}]{Kunkel2019LetME}
Kunkel, J.; Donkers, T.; Michael, L.; Barbu, C.-M.; and Ziegler, J. 2019.
\newblock Let Me Explain: Impact of Personal and Impersonal Explanations on
  Trust in Recommender Systems.
\newblock \emph{Proceedings of the 2019 CHI Conference on Human Factors in
  Computing Systems}.

\bibitem[{Lai et~al.(2023)Lai, Chen, Smith-Renner, Liao, and
  Tan}]{Lai2023TowardsAS}
Lai, V.; Chen, C.; Smith-Renner, A.; Liao, Q.~V.; and Tan, C. 2023.
\newblock Towards a Science of Human-AI Decision Making: An Overview of Design
  Space in Empirical Human-Subject Studies.
\newblock \emph{Proceedings of the 2023 ACM Conference on Fairness,
  Accountability, and Transparency}.

\bibitem[{Lai, Liu, and Tan(2020)}]{Lai2020WhyI}
Lai, V.; Liu, H.; and Tan, C. 2020.
\newblock "Why is 'Chicago' deceptive?" Towards Building Model-Driven Tutorials
  for Humans.
\newblock \emph{Proceedings of the 2020 CHI Conference on Human Factors in
  Computing Systems}.

\bibitem[{Lai and Tan(2018)}]{Lai2018OnHP}
Lai, V.; and Tan, C. 2018.
\newblock On Human Predictions with Explanations and Predictions of Machine
  Learning Models: A Case Study on Deception Detection.
\newblock \emph{Proceedings of the Conference on Fairness, Accountability, and
  Transparency}.

\bibitem[{Lee et~al.(2020)Lee, Siewiorek, Smailagic, Bernardino, and
  i~Badia}]{Lee2020CoDesignAE}
Lee, M.~H.; Siewiorek, D.~P.; Smailagic, A.; Bernardino, A.; and i~Badia, S.~B.
  2020.
\newblock Co-Design and Evaluation of an Intelligent Decision Support System
  for Stroke Rehabilitation Assessment.
\newblock \emph{Proceedings of the ACM on Human-Computer Interaction}, 4: 1 --
  27.

\bibitem[{Lee et~al.(2021)Lee, Siewiorek, Smailagic, Bernardino, and
  i~Badia}]{Lee2021AHC}
Lee, M.~H.; Siewiorek, D.~P.; Smailagic, A.; Bernardino, A.; and i~Badia, S.~B.
  2021.
\newblock A Human-AI Collaborative Approach for Clinical Decision Making on
  Rehabilitation Assessment.
\newblock \emph{Proceedings of the 2021 CHI Conference on Human Factors in
  Computing Systems}.

\bibitem[{Lee et~al.(2019)Lee, Jain, Cha, Ojha, and Kusbit}]{10.1145/3359284}
Lee, M.~K.; Jain, A.; Cha, H.~J.; Ojha, S.; and Kusbit, D. 2019.
\newblock Procedural Justice in Algorithmic Fairness: Leveraging Transparency
  and Outcome Control for Fair Algorithmic Mediation.
\newblock \emph{Proc. ACM Hum.-Comput. Interact.}, 3(CSCW).

\bibitem[{Levy et~al.(2021{\natexlab{a}})Levy, Agrawal, Satyanarayan, and
  Sontag}]{10.1145/3411764.3445522}
Levy, A.; Agrawal, M.; Satyanarayan, A.; and Sontag, D. 2021{\natexlab{a}}.
\newblock Assessing the Impact of Automated Suggestions on Decision Making:
  Domain Experts Mediate Model Errors but Take Less Initiative.
\newblock In \emph{Proceedings of the 2021 CHI Conference on Human Factors in
  Computing Systems}, CHI '21. New York, NY, USA: Association for Computing
  Machinery.
\newblock ISBN 9781450380966.

\bibitem[{Levy et~al.(2021{\natexlab{b}})Levy, Agrawal, Satyanarayan, and
  Sontag}]{Levy2021AssessingTI}
Levy, A.; Agrawal, M.; Satyanarayan, A.; and Sontag, D.~A. 2021{\natexlab{b}}.
\newblock Assessing the Impact of Automated Suggestions on Decision Making:
  Domain Experts Mediate Model Errors but Take Less Initiative.
\newblock \emph{Proceedings of the 2021 CHI Conference on Human Factors in
  Computing Systems}.

\bibitem[{Li, Lu, and Yin(2023)}]{Li_Lu_Yin_2023}
Li, Z.; Lu, Z.; and Yin, M. 2023.
\newblock Modeling Human Trust and Reliance in AI-Assisted Decision Making: A
  Markovian Approach.
\newblock \emph{Proceedings of the AAAI Conference on Artificial Intelligence},
  37(5): 6056--6064.

\bibitem[{Liu, Lai, and Tan(2021)}]{Liu2021UnderstandingTE}
Liu, H.; Lai, V.; and Tan, C. 2021.
\newblock Understanding the Effect of Out-of-distribution Examples and
  Interactive Explanations on Human-AI Decision Making.
\newblock \emph{Proceedings of the ACM on Human-Computer Interaction}, 5: 1 --
  45.

\bibitem[{Lu et~al.(2023)Lu, Li, Chiang, and Yin}]{lu2023strategic}
Lu, Z.; Li, Z.; Chiang, C.-W.; and Yin, M. 2023.
\newblock Strategic adversarial attacks in AI-assisted decision making to
  reduce human trust and reliance.
\newblock In \emph{Proceedings of the Thirty-Second International Joint
  Conference on Artificial Intelligence}, 3020--3028.

\bibitem[{Lu and Yin(2021)}]{Lu2021HumanRO}
Lu, Z.; and Yin, M. 2021.
\newblock Human Reliance on Machine Learning Models When Performance Feedback
  is Limited: Heuristics and Risks.
\newblock \emph{Proceedings of the 2021 CHI Conference on Human Factors in
  Computing Systems}.

\bibitem[{Lucic, Haned, and de~Rijke(2019)}]{Lucic2019WhyDM}
Lucic, A.; Haned, H.; and de~Rijke, M. 2019.
\newblock Why does my model fail?: contrastive local explanations for retail
  forecasting.
\newblock \emph{Proceedings of the 2020 Conference on Fairness, Accountability,
  and Transparency}.

\bibitem[{Lundberg and Lee(2017)}]{lundberg2017unified}
Lundberg, S.~M.; and Lee, S.-I. 2017.
\newblock A unified approach to interpreting model predictions.
\newblock \emph{Advances in neural information processing systems}, 30.

\bibitem[{Ma et~al.(2023)Ma, Lei, Wang, Zheng, Shi, Yin, and Ma}]{Ma2023WhoSI}
Ma, S.; Lei, Y.; Wang, X.; Zheng, C.; Shi, C.; Yin, M.; and Ma, X. 2023.
\newblock Who Should I Trust: AI or Myself? Leveraging Human and AI Correctness
  Likelihood to Promote Appropriate Trust in AI-Assisted Decision-Making.
\newblock \emph{Proceedings of the 2023 CHI Conference on Human Factors in
  Computing Systems}.

\bibitem[{Mustafatz(2023)}]{mustafaz_dia}
Mustafatz. 2023.
\newblock Diabetes Prediction Dataset.
\newblock
  \emph{https://www.kaggle.com/datasets/iammustafatz/diabetes-prediction-dataset}.

\bibitem[{Nourani et~al.(2021)Nourani, Roy, Block, Honeycutt, Rahman, Ragan,
  and Gogate}]{Nourani2021AnchoringBA}
Nourani, M.; Roy, C.; Block, J.~E.; Honeycutt, D.~R.; Rahman, T.; Ragan, E.~D.;
  and Gogate, V. 2021.
\newblock Anchoring Bias Affects Mental Model Formation and User Reliance in
  Explainable AI Systems.
\newblock \emph{26th International Conference on Intelligent User Interfaces}.

\bibitem[{Park et~al.(2019)Park, Berlin, Kirlik, and Karahalios}]{Park2019ASA}
Park, J.~S.; Berlin, R.~B.; Kirlik, A.; and Karahalios, K. 2019.
\newblock A Slow Algorithm Improves Users' Assessments of the Algorithm's
  Accuracy.
\newblock \emph{Proceedings of the ACM on Human-Computer Interaction}, 3: 1 --
  15.

\bibitem[{Passi and Vorvoreanu(2022)}]{passi2022overreliance}
Passi, S.; and Vorvoreanu, M. 2022.
\newblock Overreliance on AI Literature Review.
\newblock \emph{Microsoft Research}.

\bibitem[{Poursabzi-Sangdeh et~al.(2018)Poursabzi-Sangdeh, Goldstein, Hofman,
  Vaughan, and Wallach}]{PoursabziSangdeh2018ManipulatingAM}
Poursabzi-Sangdeh, F.; Goldstein, D.~G.; Hofman, J.~M.; Vaughan, J.~W.; and
  Wallach, H.~M. 2018.
\newblock Manipulating and Measuring Model Interpretability.
\newblock \emph{Proceedings of the 2021 CHI Conference on Human Factors in
  Computing Systems}.

\bibitem[{Pynadath, Wang, and Kamireddy(2019)}]{Pynadath2019AMM}
Pynadath, D.~V.; Wang, N.; and Kamireddy, S. 2019.
\newblock A Markovian Method for Predicting Trust Behavior in Human-Agent
  Interaction.
\newblock \emph{Proceedings of the 7th International Conference on Human-Agent
  Interaction}.

\bibitem[{Rader, Cotter, and Cho(2018)}]{Rader2018ExplanationsAM}
Rader, E.~J.; Cotter, K.; and Cho, J. 2018.
\newblock Explanations as Mechanisms for Supporting Algorithmic Transparency.
\newblock \emph{Proceedings of the 2018 CHI Conference on Human Factors in
  Computing Systems}.

\bibitem[{Ribeiro, Singh, and Guestrin(2016)}]{ribeiro2016should}
Ribeiro, M.~T.; Singh, S.; and Guestrin, C. 2016.
\newblock " Why should i trust you?" Explaining the predictions of any
  classifier.
\newblock In \emph{Proceedings of the 22nd ACM SIGKDD international conference
  on knowledge discovery and data mining}, 1135--1144.

\bibitem[{Schuff et~al.(2022)Schuff, Jacovi, Adel, Goldberg, and
  Vu}]{Schuff2022HumanIO}
Schuff, H.; Jacovi, A.; Adel, H.; Goldberg, Y.; and Vu, N.~T. 2022.
\newblock Human Interpretation of Saliency-based Explanation Over Text.
\newblock \emph{Proceedings of the 2022 ACM Conference on Fairness,
  Accountability, and Transparency}.

\bibitem[{Smith-Renner et~al.(2020)Smith-Renner, Fan, Birchfield, Wu,
  Boyd-Graber, Weld, and Findlater}]{SmithRenner2020NoEW}
Smith-Renner, A.; Fan, R.; Birchfield, M.~K.; Wu, T.~S.; Boyd-Graber, J.~L.;
  Weld, D.~S.; and Findlater, L. 2020.
\newblock No Explainability without Accountability: An Empirical Study of
  Explanations and Feedback in Interactive ML.
\newblock \emph{Proceedings of the 2020 CHI Conference on Human Factors in
  Computing Systems}.

\bibitem[{Subrahmanian and Kumar(2017)}]{subrahmanian2017predicting}
Subrahmanian, V.; and Kumar, S. 2017.
\newblock Predicting human behavior: The next frontiers.
\newblock \emph{Science}, 355(6324): 489--489.

\bibitem[{Szymanski, Millecamp, and Verbert(2021)}]{Szymanski2021VisualTO}
Szymanski, M.; Millecamp, M.; and Verbert, K. 2021.
\newblock Visual, textual or hybrid: the effect of user expertise on different
  explanations.
\newblock \emph{26th International Conference on Intelligent User Interfaces}.

\bibitem[{Tejeda et~al.(2022)Tejeda, Kumar, Smyth, and Steyvers}]{tejeda2022ai}
Tejeda, H.; Kumar, A.; Smyth, P.; and Steyvers, M. 2022.
\newblock AI-Assisted Decision-making: a Cognitive Modeling Approach to Infer
  Latent Reliance Strategies.
\newblock \emph{Computational Brain \& Behavior}, 5: 491 -- 508.

\bibitem[{Tsai et~al.(2021)Tsai, You, Gui, Kou, and
  Carroll}]{Tsai2021ExploringAP}
Tsai, C.-H.; You, Y.; Gui, X.; Kou, Y.; and Carroll, J.~M. 2021.
\newblock Exploring and Promoting Diagnostic Transparency and Explainability in
  Online Symptom Checkers.
\newblock \emph{Proceedings of the 2021 CHI Conference on Human Factors in
  Computing Systems}.

\bibitem[{van Berkel et~al.(2021)van Berkel, Gonçalves, Russo, Hosio, and
  Skov}]{Berkel2021EffectOI}
van Berkel, N.; Gonçalves, J.; Russo, D.; Hosio, S.~J.; and Skov, M.~B. 2021.
\newblock Effect of Information Presentation on Fairness Perceptions of Machine
  Learning Predictors.
\newblock \emph{Proceedings of the 2021 CHI Conference on Human Factors in
  Computing Systems}.

\bibitem[{Wang, Liang, and Yin(2023)}]{wang2023effects}
Wang, X.; Liang, C.; and Yin, M. 2023.
\newblock The effects of AI biases and explanations on human decision fairness:
  a case study of bidding in rental housing markets.
\newblock In \emph{Proceedings of the Thirty-Second International Joint
  Conference on Artificial Intelligence, IJCAI-23, Edith Elkind (Ed.).
  International Joint Conferences on Artificial Intelligence Organization},
  3076--3084.

\bibitem[{Wang, Lu, and Yin(2022)}]{wang2022will}
Wang, X.; Lu, Z.; and Yin, M. 2022.
\newblock Will you accept the ai recommendation? predicting human behavior in
  ai-assisted decision making.
\newblock In \emph{Proceedings of the ACM Web Conference 2022}, 1697--1708.

\bibitem[{Wang and Yin(2021)}]{Wang2021AreEH}
Wang, X.; and Yin, M. 2021.
\newblock Are Explanations Helpful? A Comparative Study of the Effects of
  Explanations in AI-Assisted Decision-Making.
\newblock \emph{26th International Conference on Intelligent User Interfaces}.

\bibitem[{Yang et~al.(2020)Yang, Huang, Scholtz, and Arendt}]{Yang2020HowDV}
Yang, F.; Huang, Z.; Scholtz, J.; and Arendt, D.~L. 2020.
\newblock How do visual explanations foster end users' appropriate trust in
  machine learning?
\newblock \emph{Proceedings of the 25th International Conference on Intelligent
  User Interfaces}.

\bibitem[{Yin, Vaughan, and Wallach(2019)}]{Yin2019UnderstandingTE}
Yin, M.; Vaughan, J.~W.; and Wallach, H.~M. 2019.
\newblock Understanding the Effect of Accuracy on Trust in Machine Learning
  Models.
\newblock \emph{Proceedings of the 2019 CHI Conference on Human Factors in
  Computing Systems}.

\bibitem[{Yu et~al.(2019)Yu, Berkovsky, Taib, Zhou, and Chen}]{Yu2019DoIT}
Yu, K.; Berkovsky, S.; Taib, R.; Zhou, J.; and Chen, F. 2019.
\newblock Do I trust my machine teammate?: an investigation from perception to
  decision.
\newblock \emph{Proceedings of the 24th International Conference on Intelligent
  User Interfaces}.

\bibitem[{Zhang, Liao, and Bellamy(2020)}]{Zhang2020EffectOC}
Zhang, Y.; Liao, Q.~V.; and Bellamy, R. K.~E. 2020.
\newblock Effect of confidence and explanation on accuracy and trust
  calibration in AI-assisted decision making.
\newblock \emph{Proceedings of the 2020 Conference on Fairness, Accountability,
  and Transparency}.

\end{thebibliography}

\newpage
\appendix
\section{Literature Review}
We screened research papers related to AI-assisted decision making that are published between 2018 and 2021 in the ACM CHI Conference on Human Factors in Computing Systems (CHI), ACM Conference on Computer-supported Cooperative Work and
Social Computing (CSCW), ACM Conference on Fairness, Accountability, and Transparency (FAccT), and ACM Conference on Intelligent User Interfaces (IUI) to identify different forms of AI assistance developed in the literature. We grouped different forms of AI assistance into a few categories: 
\begin{enumerate}
    \item \emph{Immediate assistance}~\cite{Lai2018OnHP,Liu2021UnderstandingTE,Nourani2021AnchoringBA,Green2019ThePA,Tsai2021ExploringAP,Bansal2020DoesTW,Buccinca2021ToTO,Feng2018WhatCA,Guo2019VisualizingUA,Lee2020CoDesignAE,Lee2021AHC,Levy2021AssessingTI,Cheng2019ExplainingDA,Lai2020WhyI,PoursabziSangdeh2018ManipulatingAM,Chromik2021ITI,Jacobs2021DesigningAF,SmithRenner2020NoEW,Desmond2021IncreasingTS,Buccinca2020ProxyTA,Gajos2022DoPE,Gomez2020ViCEVC,Abdul2020COGAMMA,Brown2019TowardAA, Cai2019TheEO, Buccinca2020ProxyTA,Szymanski2021VisualTO,Green2019DisparateIA,DeArteaga2020ACF,Yang2020HowDV,Kunkel2019LetME,Das2020LeveragingRT,Yu2019DoIT,10.1145/3359284,Harrison2020AnES,Kocielnik2019WillYA} 

    \item \emph{Delayed recommendation}~\cite{Zhang2020EffectOC,Dodge2019ExplainingMA,Wang2021AreEH,Yin2019UnderstandingTE,Buccinca2021ToTO,Lu2021HumanRO,PoursabziSangdeh2018ManipulatingAM,10.1145/3359280,Park2019ASA}

    \item \emph{Explanation only}~\cite{Lai2018OnHP,Alqaraawi2020EvaluatingSM,Lucic2019WhyDM,Rader2018ExplanationsAM,Berkel2021EffectOI,Buccinca2020ProxyTA,Gajos2022DoPE,Anik2021DataCentricEE,Lucic2019WhyDM,Rader2018ExplanationsAM}

    \item \emph{Interaction between human and AI}: Different from the three ``static'' types of AI assistance, this form of AI assistance emphasizes the interaction between human decision maker (DM) and the AI assistant. For example, during the collaboration with AI, AI can provide the accuracy feedback to help DMs recalibrate their trust in AI~\cite{Bansal2020DoesTW,Yu2019DoIT}. In addition, DMs may actively explore the decision space of AI assistants~\cite{10.1145/3290605.3300234,10.1145/3411764.3445522}, or they can be provided with interactive explanations to gain a deeper understanding of how AI models arrive at their decisions~\cite{10.1145/3290605.3300234, Yang2020HowDV, SmithRenner2020NoEW,Liu2021UnderstandingTE,Cai2019HumanCenteredTF}, thereby enhancing their appropriate trust in AI assistants.

\end{enumerate}
Given the limited number of papers in the \emph{Interaction between Human and AI} category, and their unique interaction designs, in this study, we focus on building computational framework to model how the first three types of AI assistance influence human DMs.

\section{Additional Details of Human-Subject Experiment}
\noindent \textbf{Data Validity Check.} To verify the engagement of subjects in our study, an attention check question was included in which subjects were instructed to select a pre-specified option. Among the 285 workers participated in our study, 202 passed the attention check question. Only the data from them were considered as valid and used to train/evaluate our models. Also, as an evidence of ``consistency'', across all decision making tasks, the average fraction of subjects who agreed with the majority decision on the task was 82\% (though decision makers did not need to agree with others' decisions).

\vspace{2pt}
\noindent \textbf{Working Time.} The mean  completion times for a decision making task and their standard deviations in different treatments are: {\em Independent}: $4.61 \text{s} \, \pm 3.27 \text{s}$, {\em Immediate assistance}: $5.03 \text{s} \, \pm 3.42 \text{s}$, {\em Delayed recommendation}:  $9.89 \text{s} \, \pm 6.07 \text{s}$, {\em Explanation only}:   $5.45 \text{s} \, \pm 3.54 \text{s}$. 

\begin{table}[t]

\centering
\resizebox{\linewidth}{!}{
\begin{tabular}{cccccc}
\hline
Number of Training Instances & 5     & 10   & 15    & 20    & 25    \\ \hline
Deterministic Decision Model & 0.514      & 0.469    & 0.454      & 0.434      &  0.416    \\
Ours                         & 0.430 & 0.422 & 0.413 & 0.402 & 0.394 \\ \hline
\end{tabular}
}
\caption{Comparing  the performance of our method against an alternative that substitutes the distribution of decision model $q_{\phi}(\bm{w}_h)$ of our method  with a deterministic logistic regression model in the \emph{Delayed recommendation} scenario. NLL is adopted as the evaluation metric, with a lower NLL denoting superior performance.
}
\label{tab:ablation}
\end{table}

\section{Ablation Study}
In our approach, we adopt a probabilistic framework to learn a distribution of the independent human decision model $q_{\phi}(\bm{w}_h)$. In this study, we conducted an ablation study by replacing the distribution of the decision model $q_{\phi}(\bm{w}_h)$ with a deterministic logistic regression model that can be learned in the \emph{Delayed recommendation} scenario (because human DMs need to first provide their initial decision before the AI recommendation is revealed). 
As shown in Table~\ref{tab:ablation}, we observed that our approach consistently outperforms the counterpart using the deterministic decision model as we vary the number of training instances.

\section{The Potential Influence of the LLM-Powered Decision Aids on Humans}
The AI model we used in our study was a supervised learning model that was trained independently without human feedback. However, with the rapid development of large language models (LLMs), one may envision that future AI-based decision aids can be powered by LLMs. It is known that LLMs may learn from human feedback and may have the tendency to 
provide affirmative responses to humans, which could reinforce human DMs' beliefs and biases in the long run. 
This could be particularly concerning if the DM is intentionally providing feedback to LLMs in a way that seeks approval for a decision that is flawed or biased. As the LLM keeps internalizing the human DM's biases through their feedback and learns to provide affirmative response to DMs, the DM might perceive the AI's affirmative response as an endorsement from an expert, leading to an increased likelihood of confirmation bias. Moreover, the consistent affirmative feedback from LLMs could subtly alter the human cognitive decision making process. For example, if LLMs continually affirm DMs' decisions or ideas, it may lead to DMs' overconfidence in their decisions. Developing computational frameworks to characterize the dynamics between the influence of AI assistance to human DMs and the influence of human DMs' feedback to AI assistance for future AI-based decision aids that are powered by LLMs can be a very interesting future direction.

\end{document}